\documentclass[
    prb,
    twoside,
    twocolumn,
    10pt,
    floatfix,
    showpacs,
    citeautoscript,
    superscriptaddress,
]{revtex4-2}

\usepackage{amsmath}
\usepackage{amssymb}
\usepackage{glossaries}
\usepackage{graphicx}
\usepackage[version=4]{mhchem}
\usepackage{siunitx}
\usepackage[colorlinks=true]{hyperref}

\newcommand \Title{
    Tuning the through-plane lattice thermal conductivity\texorpdfstring{\\}{}
    in van-der-Waals structures through rotational (dis)ordering
}

\hypersetup{
    pdfauthor={F. Eriksson, E. Fransson, C. Linderälv, Z. Fan, and P. Erhart},
    pdftitle={\Title},
    pdffitwindow=false,
    pdfstartview={FitH},
    pdfnewwindow=true,
    colorlinks=true,
    linkcolor=black,
    citecolor=black,
    filecolor=black,
    urlcolor=black,
}

\setacronymstyle{long-short}
\newacronym{bop}{BOP}{bond order potential}
\newacronym{cx}{vdW-DF-cx}{van-der-Waals density functional with consistent exchange}
\newacronym{dft}{DFT}{density functional theory}
\newacronym{emd}{EMD}{equilibrium molecular dynamics}
\newacronym{gk}{GK}{Green-Kubo}
\newacronym{la}{LA}{longitudinal acoustic}
\newacronym{lo}{LO}{longitudinal optic}
\newacronym{ltc}{LTC}{lattice thermal conductivity}
\newacronym{md}{MD}{molecular dynamics}
\newacronym{mlp}{MLP}{machine-learned potential}
\newacronym{nep}{NEP}{neuroevolution potential}
\newacronym{pbte}{PBTE}{Peierls-Boltzmann transport equation}
\newacronym{scan}{SCAN}{strongly constrained and appropriately normed}
\newacronym{ta}{TA}{transverse acoustic}
\newacronym{to}{TO}{transverse optic}
\newacronym{vdw}{vdW}{van-der-Waals}
\newacronym{xc}{XC}{exchange-correlation}

\renewcommand{\vec}[1]{\boldsymbol{#1}}
\DeclareSIUnit\angstrom{\text{\AA}}

\begin{document}

\title{\Title}

\newcommand{\addchalmers}{Department of Physics, Chalmers University of Technology, SE-41296, Gothenburg, Sweden}
\newcommand{\addbohai}{College of Physical Science and Technology, Bohai University, Jinzhou 121013, P. R. China}

\author{Fredrik Eriksson}
\affiliation{\addchalmers}
\author{Erik Fransson}
\affiliation{\addchalmers}
\author{Christopher Linderälv}
\affiliation{\addchalmers}
\author{Zheyong Fan}
\affiliation{\addbohai}
\author{Paul Erhart}
\affiliation{\addchalmers}
\email{erhart@chalmers.se}


\begin{abstract}
It has recently been demonstrated that \ce{MoS2} with irregular interlayer rotations can achieve an extreme anisotropy in the lattice thermal conductivity (LTC), which is for example of interest for applications in waste heat management in integrated circuits.
Here, we show by atomic scale simulations based on machine-learned potentials that this principle extends to other two-dimensional materials including C and \ce{BN}.
In all three materials introducing \emph{rotational disorder} drives the through-plane LTC to the glass limit, while the in-plane LTC remains almost unchanged compared to the ideal bulk materials.
We demonstrate that the ultralow through-plane LTC is connected to the collapse of their transverse acoustic modes in the through-plane direction.
Furthermore, we find that the twist angle in periodic moiré structures representing \emph{rotational order} provides an efficient means for tuning the through-plane LTC that operates for all chemistries considered here.
The minimal through-plane LTC is obtained for angles between 1 and \SI{4}{\degree} depending on the material, with the biggest effect in \ce{MoS2}.
The angular dependence is correlated with the degree of stacking disorder in the materials, which in turn is connected to the slip surface.
This provides a simple descriptor for predicting the optimal conditions at which the LTC is expected to become minimal.
\end{abstract}

\maketitle

\section{Introduction}
Understanding the atomic scale dynamics of materials is important from both conceptual and practical vantage points.
They are not only fundamental to the thermodynamic and kinetic properties of materials but also strongly affect electronic transport and optical response.
The \gls{ltc} in particular is important for applications in, e.g., thermoelectrics and thermal management \cite{Rowe2006}.
In the latter case, anisotropic thermal conductors have been proposed as an efficient means for removing waste heat \cite{minnich_exploring_2016, cui_emerging_2020, moore_emerging_2014}.

\begin{figure*}
    \centering
    \includegraphics[width=0.98\textwidth]{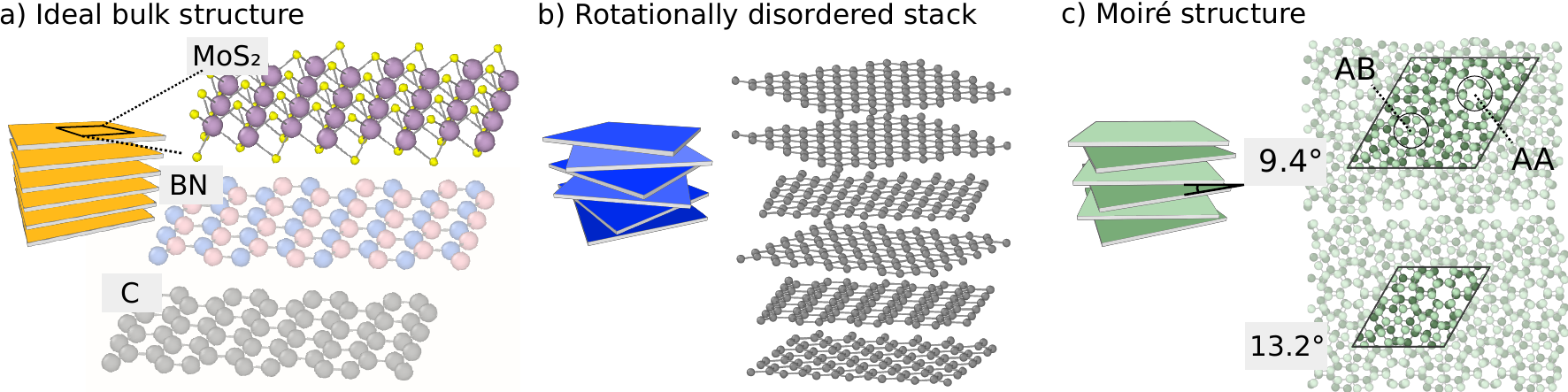}
    \caption{
        \Acrlong{vdw} structures consist of monolayers with strong intralayer and weak (\gls{vdw}-mediated) interlayer interactions.
        a) Ideal bulk structures of \ce{MoS2}, \ce{BN} and \ce{C} (graphite) are characterized by perfect registry between the layers (only monolayers shown).
        b) In rotationally disordered stacks the twist angles between the monolayers are random.
        c) In (bulk) moiré structures every other layer is rotated with the same twist-angle.
        They can serve as simple model systems providing insight into the mechanisms giving rise to ultralow through-plane \gls{ltc} and large anisotropy.
    }
    \label{fig:overview}
\end{figure*}

\Gls{vdw} materials consist of quasi two-dimensional layers with strong intralayer and weak (\gls{vdw}-mediated) interlayer interactions.
In the ideal bulk form of, e.g., \ce{MoS2}, \ce{C} (graphite) or \ce{BN}, the layers are highly ordered, typically with a two-layer repetition period (\autoref{fig:overview}a).
In disordered \gls{vdw} materials on the other hand the orientation (i.e., the rotational angle) between the layers is random (\autoref{fig:overview}b).
Such materials have naturally high anisotropy ratios, a property that is of potential interest, especially for thermal management applications \cite{norley_development_2001, chung_performance_2012, chiritescu_ultralow_2007, ChenSooPop19, VazYalMun19}.
Yet the artificial synthesis of materials with comparable anisotropies and through-plane conductivities of less than \SI{0.1}{\watt\per\meter\per\kelvin} was only accomplished recently \cite{Kim2021}.
This progress has been enabled by new synthesis routes that allow manipulation of the angles between individual layers in many-layer samples \cite{kim_tunable_2017}.

It is well known that interlayer rotations in two-dimensional \gls{vdw}-bonded structures lead to the emergence of moiré patterns (\autoref{fig:overview}c) and novel properties \cite{Choi2019, Haddadi2020, Xian2019, Lian2019, Wu2019, Regan2020, Naik2022}.
The twist angle provides an additional (structural) degree of freedom that can be used, for example, to induce superconductivity in bilayer graphene \cite{Cao2018, Cao2018b}.
Given the effect of the twist angle on electronic properties it is natural to ask whether it can also be used to manipulate the \gls{ltc} of these materials.
If the goal is to maximize the \gls{ltc} anisotropy, lowering the through-plane \gls{ltc} is key as the the in-plane \gls{ltc} is bounded from above by the \gls{ltc} of the corresponding monolayer.

Several mechanisms may play a role in lowering the through-plane \gls{ltc} of \gls{vdw} structures in general \cite{ChenSooPop19, WeiCheDam13} including interlayer rotations \cite{NieZhaDen19, OuyQinUrb20, ChoDemChe22, SunHuZha22}.
Interlayer rotations cause the atoms in adjacent layers to be pushed out of registry.
This drastically reduces the shear resistance and is manifested in the localization of the corresponding \gls{ta} phonon modes \cite{ErhHylLin15, Maity2020, Kim2021, SunHuZha22}.
Moreover, with decreasing twist angle the moiré cell grows, leading to more extended displacement patterns.
At the same time, there is a limit to the disorder associated with these displacements since for sufficiently small angles the layers reconstruct into regions of the energetically favored bulk stacking that are separated by domain walls \cite{Carr2018, Yoo2019, McGilly2020, Quan2021, Naik2022}.
This reconstruction is governed by the intrinsic properties of the material, such as the elastic constants and the interlayer potential energy landscape.
The interplay of these factors can be expected to lead to a minimum in the through-plane \gls{ltc} as a function of twist angle.
Quantitative assessments of these effects require, however, accurate and predictive atomic-scale simulations that can guide future experimental studies.

For materials with relatively high symmetry and modest unit cell sizes the \gls{ltc} can be accurately predicted and analyzed in the framework of the \gls{pbte} using force constants calculated via electronic structure methods such as \gls{dft}.
Due to the scaling of both the \gls{pbte} and the electronic structure calculations this approach becomes, however, prohibitive for materials with larger unit cells and/or lower symmetry.
This challenge can be overcome using \gls{gk} methods in conjunction with \gls{md} simulations, which, however, require suitable interatomic potentials.

Here, we employ the \gls{gk} approach in combination with \glspl{mlp} to analyze the \gls{ltc} for three prototypical \gls{vdw} materials with interlayer rotations: graphite/graphene (C), hexagonal boron-nitride (BN), and molybdenum disulfide (\ce{MoS2}).
We focus on two types of three-dimensionally periodic structures that are compared to the ideal bulk structures (\autoref{fig:overview}a):
(1) \emph{stacks} with arbitrary rotation angles and small in-plane strains comprising up to 10 layers per unit cell, representing \emph{rotational disorder} (\autoref{fig:overview}b) and (2) \emph{moiré structures} with a single rotation angle, i.e., the primitive cell contains two monolayers with a specific rotation angle, representing \emph{rotational order} (\autoref{fig:overview}c).
We show that for all three materials rotational disorder gives rise to a systematic and substantial reduction in the through-plane \gls{ltc} without strongly affecting the in-plane conductivity.
In all cases, we find that the stacks display glass-like conduction with the largest \gls{ltc} anisotropy appearing in C, for which we obtain a ratio of over \num{1000} at room temperature.

Further insight is provided by the dependence of the \gls{ltc} on the twist angle in periodic moiré structures, which we relate to the atomic level reconstructions.
The latter connection enables a particular simple interpretation of the angular dependence of the \gls{ltc} in terms of the slip surfaces of the different materials.
Our results demonstrate that rotational disorder can be used for manipulating the \gls{ltc} in layered materials that is largely agnostic to chemistry, and provide insight into the underlying mechanisms.
We expect that these insights can be exploited, e.g., for developing materials for heat management in integrated circuits, and more generally contribute to not only understanding but also controlling thermal conduction at the nanoscale.

\section{Results and discussion}

\begin{figure*}
\centering
\includegraphics[width=0.98\textwidth]{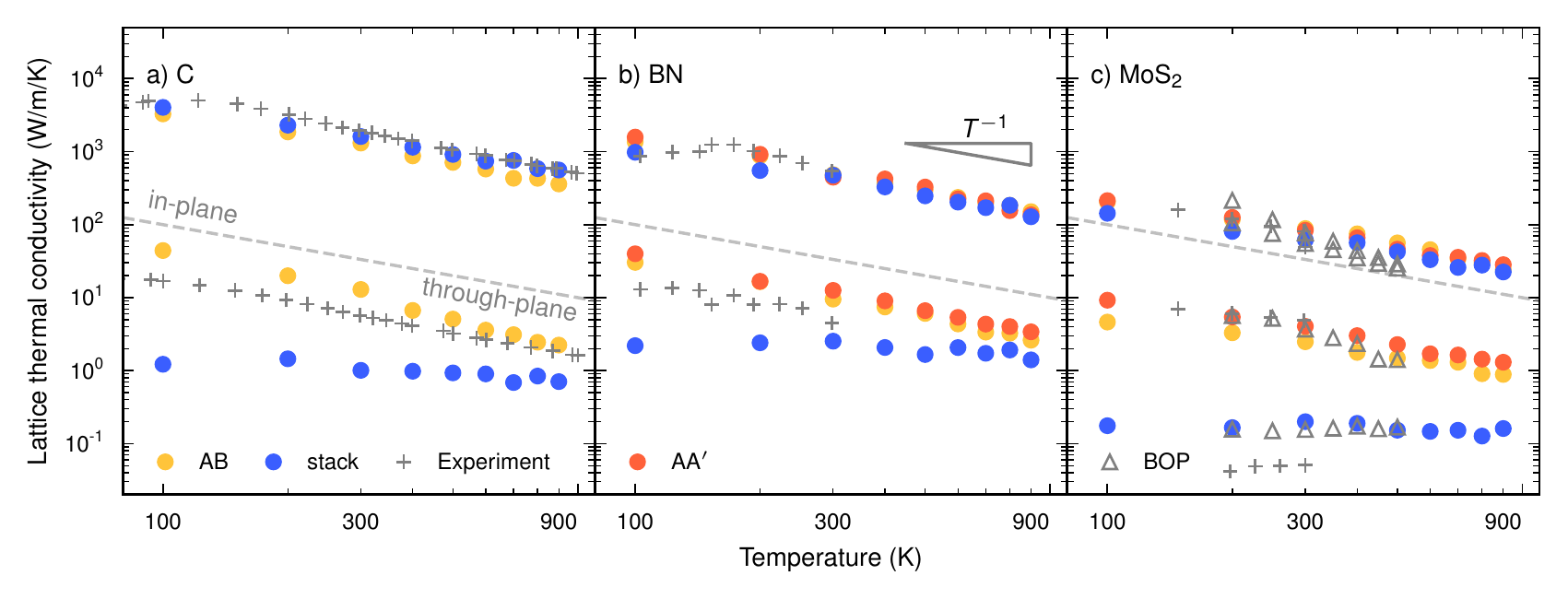}
\caption{
    \Gls{ltc} for (a) C, (b) \ce{BN} and (c) \ce{MoS2} as a function of temperature for the ideal bulk systems as well as a rotationally disordered stack system with random interlayer rotations.
    The dashed gray line separates the in-plane and out-of-plane components of the \gls{ltc} tensor.
    The gray plus signs indicate experimental data from Ref.~\citenum{Ho1972} (C; graphite), Ref.~\citenum{yuan_modulating_2019} (\ce{BN}), and Ref.~\citenum{Kim2021} (\ce{MoS2}).
    The triangles in panel (c) represent results from simulations based on a \gls{bop} model from Ref.~\citenum{Kim2021}.
    The statistical errors for the thermal conductivity are about the size of the markers across all data points.
}
\label{fig:kappa_stack}
\end{figure*}

\subsection{\texorpdfstring{\Gls{ltc}}{LTC} in bulk and disordered stacks}

\paragraph*{Carbon.}
To begin with we consider the temperature dependence of the \gls{ltc} for the stack structure and compare it with the ideal bulk structures for the case of carbon (\autoref{fig:kappa_stack}).
For the in-plane \gls{ltc} of the ideal bulk structure (AB, graphite) the simulations are in very good agreement with experimental data. \cite{Ho1972}
This applies not only for the \gls{mlp} based on \gls{cx} shown here but also for models based on the PBE+D3 and \gls{scan} \gls{xc} functionals, as shown by \gls{pbte} calculations (\autoref{sfig:bte}).
For the through-plane \gls{ltc} the simulations somewhat overestimate the experimental data for temperatures below approximately \SI{600}{\kelvin}.
This is expected as the through-plane \gls{ltc} is not only more difficult to measure but also much more sensitive to sample purity and (small) variations in the aspect ratio.
This is also evident from the comparison with the \gls{pbte} results for the models based on other \gls{xc}, which overestimate the aspect ratio and underestimate the through-plane \gls{ltc} (\autoref{sfig:bte}).

The \gls{cx} method and accordingly the \gls{cx}-based \gls{mlp} achieve good overall agreement with the structural parameters as well as experimental data, demonstrating that they capture the vibrational excitations that govern thermal conduction in this material.

Moving on to the stack system with rotational disorder, one observes a substantial drop in the through-plane \gls{ltc} while the in-plane \gls{ltc} remains at the level of the ideal bulk system.
For example, at \SI{300}{\kelvin} the through-plane \gls{ltc} is reduced by more than a factor of ten, leading an anisotropy ratio between the fast and the slow \gls{ltc} components of more than \num{1000}.
The \gls{ltc} is moreover constant over the temperature range considered here, a behavior commonly observed in glasses. \footnote{
    Since our simulations are classical the \gls{ltc} remains constant even at low temperatures.
    If quantum effects were included one would expect a drop of the \gls{ltc} at low temperatures.
}
As further discussed below, this can be understood as the phonon mean free path for through-plane transport being approximately limited to the interlayer distance.

It is noteworthy that the in-plane \gls{ltc} for the stack even exceeds that of graphite.
This effect can be attributed to the weaker coupling between layers, which affects the flexural modes and thereby the \gls{ltc}.
This effect is also apparent in the larger in-plane \gls{ltc} of graphene sheets compared to graphite \cite{Balandin2008, Ghosh2008, Lindsay2010, CheFanZha23}.

\paragraph*{Boron nitride.}
In \ce{BN} the behavior of the thermal conductivities is qualitatively the same as for C (\autoref{fig:kappa_stack}b).
While there are two types of ideal bulk stackings, AA$^\prime$ and AB, the difference in \gls{ltc} between these two structures is minimal.
The agreement with experimental data \cite{yuan_modulating_2019} for the in-plane conductivity is very good and the \gls{ltc} falls off with $T^{-1}$.
For the through-plane conductivity the simulations yield slightly higher conductivities than experiment, equivalent and for similar reasons as in the the case of C.
Also the behavior of the \gls{ltc} for the stack system is similar, showing the same kind of temperature independent conductivity.
The reduction in the through-plane \gls{ltc} when going from the ideal to the stack system is, however, notably smaller than in the case of C, leading only an anisotropy ratio of about \num{200} at \SI{300}{\kelvin}.

\paragraph*{Molybdenum disulfide.}
For the ideal bulk structure of \ce{MoS2} both the in-plane and through-plane conductivity are in very good agreement with experimental data (\autoref{fig:kappa_stack}c) \cite{Kim2021}, and, as in the case of \ce{BN}, the \gls{ltc} is practically the same for AA$^\prime$ and AB structures.
As for the other two materials, the in-plane \gls{ltc} for the stack system is almost unchanged compared to the ideal bulk system, while the through-plane \gls{ltc} exhibits a glass-like temperature dependence, achieving an anisotropy ratio of about \num{300} at \SI{300}{\kelvin}.

\begin{figure*}
\centering
\includegraphics{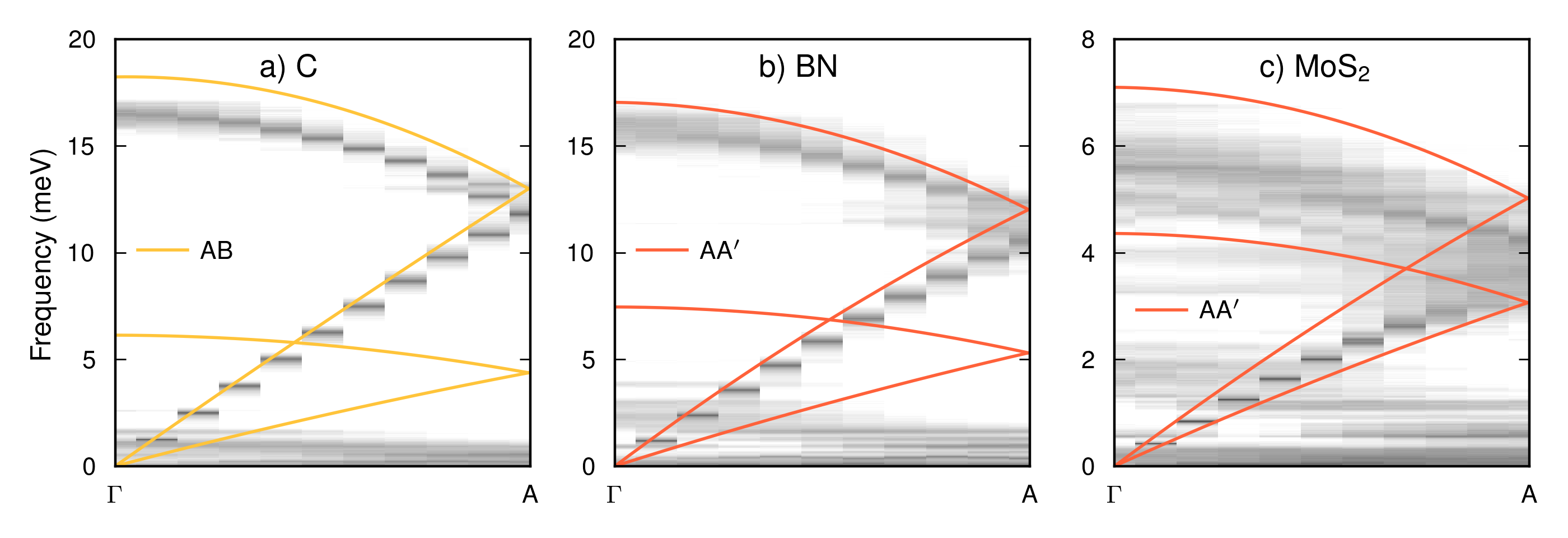}
\caption{
    Phonon dispersions obtained via mode projections from \gls{md} simulations for ideal bulk (lines) and stack structures (heat map) at \SI{300}{\kelvin} for (a) carbon, (b) boron nitride, and (c) molybdenum disulfide along the $\Gamma \to$A direction ($[0,0,1]$).
    The heat maps show the natural logarithm of the velocity power spectra obtained by projection onto the normal modes of the respective ideal bulk structure (see \autoref{sect:computational-details} for details).
}
\label{fig:phonon-dispersions}
\end{figure*}

The calculated through-plane \gls{ltc} for the stack is notably higher compared to experiments \cite{Kim2021}.
This is likely due to other effects, besides the stacking, being at play in the experimental study that are not captured in the simulations, including, e.g., the presence of defects \cite{GabSurFar20, PolPanBer20} and the contribution of interface resistivity in the experimental devices.

Here, we also include a comparison with \gls{ltc} data obtained previously \cite{Kim2021} via \gls{md} simulations using a \gls{bop} model \cite{liang_parametrization_2009, stewart_atomistic_2013}.
While the latter yields a somewhat steeper temperature dependence for the in-plane conductivity, the results are overall very close, including in particular the through-plane \gls{ltc} for the stack system.
This agreement is remarkable given that the \gls{nep} models used in the present work and the \gls{bop} model employ very different functional forms and were constructed using different reference data and design principles.
This goes to show that the effect revealed here is not sensitive to the specifics of the underlying model but rather an intrinsic feature of material and structure.

\subsection{Rotational disorder in the phonon dispersion}

In order for a mode to contribute to conduction in the through-plane direction, it must have a non-zero group velocity component in the $z$-direction, which applies for modes that fall within a rather narrow cone along the $\Gamma$--A path \cite{Gu19}.
To reveal the microscopic mechanisms that lead to the dramatic reduction in the through-plane \gls{ltc} in the stack structures, it is therefore instructive to inspect the vibrational spectra along $\Gamma$--A.
This analysis (see \autoref{sect:mode-projection} in the Methods section below) reveals that the dispersion of the \gls{la} modes is only very weakly affected when rotational disorder is introduced (\autoref{fig:phonon-dispersions}).
At the same time one observes a collapse of the \gls{ta} and the lowermost \gls{to} modes in all stack systems.
In other words, these modes soften significantly and the frequencies become nearly independent of the momentum vector, as previously shown \cite{Kim2021}  in the case of \ce{MoS2} using a \gls{bop} model \cite{liang_parametrization_2009}.

One can understand the collapse of the \gls{ta} modes as being related to a large reduction of the shear resistance.
The latter arises because the interlayer rotations push the layers out of registry, which reduces the energy barriers that need to be overcome to shear neighboring layers relative to each other.

A further important observation is that the phonon lifetimes of the \gls{la} (and the \gls{lo}) modes can drop by as much as two orders of magnitude when going from the ideal bulk to the stack structures (\autoref{sfig:LALO_lifetimes}).
Interestingly one can show that reducing the frequencies of the \gls{ta} modes while leaving the phonon-phonon interaction (i.e., the third-order force constants) unchanged, is sufficient for achieving a dramatic drop in the through-plane conductivity \emph{without} introducing explicit disorder (\autoref{snote:fcp_bte} and \autoref{sfig:fcp-bte-phonons}).
This demonstration is distinct from the observation that a rescaling of the (entire) interaction potential leads to a negative correlation between the in-plane and through-plane conductivities \cite{WeiCheDam13}.

\subsection{\texorpdfstring{\Gls{ltc}}{LTC} reduction in terms of phonon scattering}

To obtain a conceptually intuitive understanding of the low through-plane \gls{ltc} in rotationally disordered systems, recall that according to the linearized solution of the \gls{pbte} \cite{Zim60} the \gls{ltc} is given by a summation over all phonon modes
\begin{align}
    \mathbf{\kappa}
    =
    \frac{1}{N_{\vec{q}} \Omega} \sum_{\vec{q}j} \tau_{\vec{q}j} \vec{v}_{\vec{q}j} \otimes \vec{v}_{\vec{q}j} c_{\vec{q}j}.
    \label{eq:kappa}
\end{align}
Here, $\tau_{\vec{q}j}$ is the lifetime, $\vec{v}_{\vec{q}j}$ is the group velocity, $c_{\vec{q}j}$ is the mode specific heat capacity, which in the classical limit is a constant, while $\vec{q}$ and $j$ indicate phonon momentum and branch.
Finally, $N_{\vec{q}}$ is the number of $\vec{q}$-points included in the summation and $\Omega$ is the unit cell volume.

In this picture, a drop in the \gls{ltc} can thus result from a reduction of the lifetimes or group velocities.
As the group velocity of the heat carrying longitudinal modes is only weakly affected by interlayer rotations (\autoref{fig:phonon-dispersions}), the majority of the reduction must be attributable to a reduction of the lifetimes, which is consistent with our analysis (\autoref{sfig:LALO_lifetimes}).

Due to the large anisotropy in \gls{vdw} structures, the phonon modes in these materials can be separated into two distinct regions in the Brillouin zone, predominantly contributing to the in-plane and out-of-plane \gls{ltc}, respectively \cite{Gu19}.
In the former region the group velocities are close to zero in the through-plane direction and the modes are monolayer-like.
The modes in the second set are confined to a narrow cone along $\Gamma$-A with in-plane group velocities that are close to zero contributing very little to the in-plane thermal conductivity.
As a result of this separation, the collapse of the \gls{ta} mode in  rotationally disordered stacks and the decrease of the \gls{la} mode lifetimes have almost no impact on the in-plane \gls{ltc}.

\subsection{\texorpdfstring{\Gls{ltc}}{LTC} in rotationally ordered systems}

\begin{figure}
\centering
\includegraphics{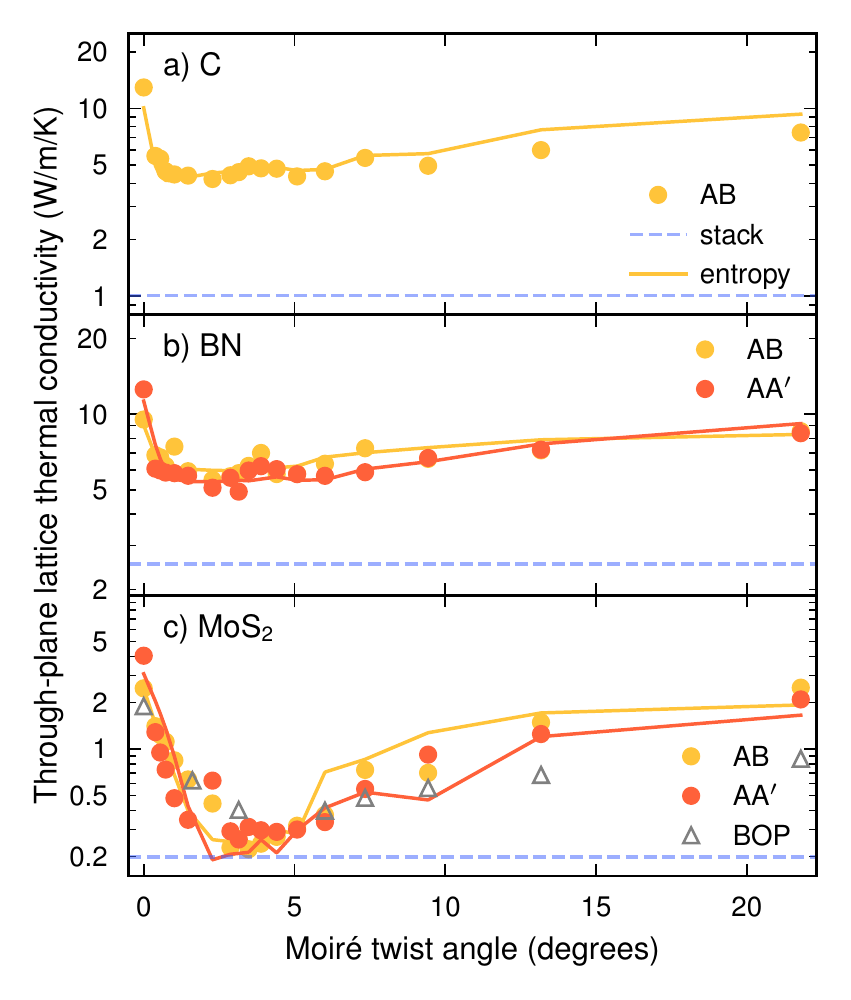}
\caption{
    \Glspl{ltc} of moiré structures as a function of twist angle at \SI{300}{\kelvin} in (a) carbon, (b) \ce{BN}, and (c) \ce{MoS2}.
    The statistical errors for the thermal conductivity is about twice as large as the markers across all data points.
    The negative entropy $-S$ of the stacking order parameter, see Eq.~\eqref{eq:entropy}, is shown as solid lines using an arbitrary $y$-scale, demonstrating the correlation between stacking disorder and low \gls{ltc}.
    The dashed lines indicate the \gls{ltc} of the stack systems (compare \autoref{fig:kappa_stack}).
}
\label{fig:ltc-vs-angle}
\end{figure}

To gain further insight into the reduction of the through-plane \gls{ltc} in \emph{rotationally disordered} stacks of layers, it is useful to study the dependence of the \gls{ltc} on the rotation angle in \emph{rotationally ordered} systems.
For all three materials and both stackings we observe that there exists a minimum in the \gls{ltc} between 0 and \SI{5}{\degree} (\autoref{fig:ltc-vs-angle}).
In \ce{MoS2} the minimum is very pronounced and for twist angles around \SI{3}{\degree} the through-plan \gls{ltc} approaches the same value as for the stack.
For C and \ce{BN} the minima are wider and less pronounced, and the minimal \gls{ltc} values are still notably above the values obtained in the respective stacks.

The theoretical lower limit for the \gls{ltc} in dense materials is reached when the mean free path available of the heat carrying phonon modes becomes comparable to the interatomic distances \cite{Cahill1992}.
For example, in the case of turbostratically deposited \ce{MoS2}, which achieves an ultralow \gls{ltc} in the through-plane direction \cite{chiritescu_ultralow_2007}, the effective mean free path approaches the interlayer spacing \cite{ErhHylLin15}.
While the layer spacing in \ce{MoS2} is about \SI{6}{\angstrom} it only amounts to about \SI{3}{\angstrom} in C and \ce{BN}.
This suggests that while interlayer rotations with a periodicity of two layers are sufficient to approach the minimal mean free path, lower sequences are required to achieve the same effect in C and \ce{BN}.

Lastly, it is striking that the through-plane \gls{ltc} in \ce{MoS2} obtained with the \gls{nep} is in close agreement with the results obtained using a \gls{bop} model both in terms of the absolute values and the position of the angle corresponding to the minimal \gls{ltc}.
This provides a further indication that the results obtained here are caused by generic microscopic mechanism rather than tied to the details of the atomic interaction models.

\subsection{Reconstruction in moiré structures}

The prediction of the \gls{ltc} from atomic scale simulations is computational demanding.
It is therefore desirable to identify simpler predictors for the observed behavior.
At the first level of abstraction, one can consider the atomic displacement patterns that emerge in the moiré structures.

The interlayer rotations force the atoms in neighboring layers into energetically less favorable stackings (\autoref{sect:stackings}).
To minimize the energy the atoms in each layer then undergo displacements, which gives rise to a reconstruction into regions that are similar to the ideal bulk stackings separated by ``domain walls''.
The size of each of these regions depends on the material specific energy landscape (\autoref{sect:slip-surface}).

\begin{figure*}
\centering
\includegraphics{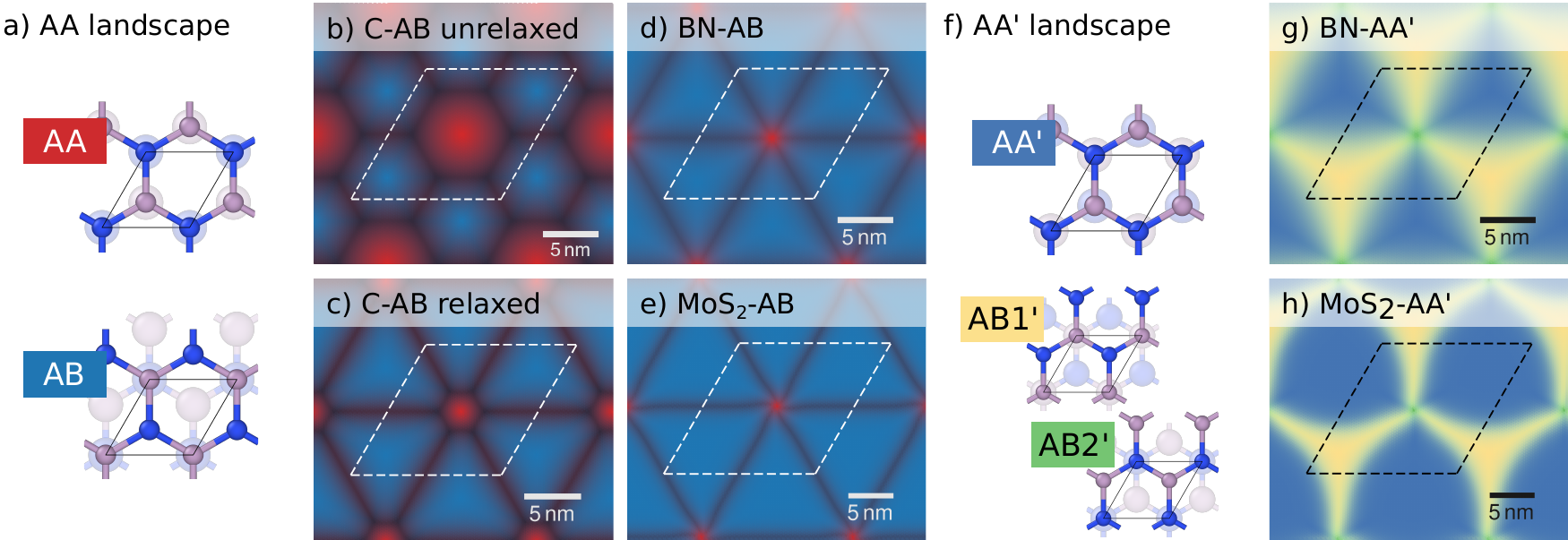}
\caption{
    Variation of the local environment with position in a moiré structure with a twist angle of \qty{1.02}{\degree} for (a--e) AB stacking and (f--h) AA$^\prime$ stacking.
    The different colors indicate the similarity with different (ideal) bulk stacking sequences shown in (a,f) obtained via template matching (see \autoref{sect:stacking-order-parameter} in the Methods section).
    Comparison of relaxed (c) and unrelaxed structures (d) shows how reconstruction allows the system to form extended regions of energetically more favorable stacking sequences.
    The eventual structure is the result of a balance between in-plane strain and domain wall formation.
    The size of the different regions in the different materials correlates with their respective slip surfaces (\autoref{fig:slip-surface}).
}
\label{fig:moire-template-matching}
\end{figure*}

To quantify the size of these different regions we can define a simple order parameter (see Eq.~\eqref{eq:order-parameter}).
For the AB-based moiré structures this reveals extended AB regions separated by domain walls with small AA regions at the domain wall intersections \cite{CazClaEng23}.
The AA$^\prime$-based structures on the other hand feature extended AA$^\prime$ domains separated by large AB1$^\prime$ domains and small AB2$^\prime$ regions at the intersections.
The regions with the ideal bulk stacking are most extended for \ce{MoS2} compared to C and \ce{BN} (\autoref{fig:moire-template-matching}).

Using the order parameter $\alpha_i$ defined in Eq.~\eqref{eq:order-parameter}, we can moreover define a measure for the stacking disorder by estimating the entropy of the probability distribution over the order parameters $\alpha$ as
\begin{align}
    S = \sum_i p_i(\alpha) \ln p_i(\alpha).
    \label{eq:entropy}
\end{align}
Here, $p_i(\alpha)$ is the probability distribution over $\alpha$ found in a structure.
The entropy $S$ thus measures the relative occurrence of different local stackings in the system.

As shown by the solid lines in \autoref{fig:ltc-vs-angle}, the negative entropy $-S$ exhibits a very similar angular dependence as the through-plane \gls{ltc}.
This indicates that the \gls{ltc} is to a large extent correlated with the disorder in the system and conversely that $S$ can serve as a simple (and much cheaper) indicator for the angular dependence of the conductivity.

\subsection{Slip surface}
\label{sect:slip-surface}

It is now natural to ask which materials parameters determine the reconstruction in the moiré structures.
The latter are driven by the energy gain when forming regions that conform to the low energy bulk stacking, which needs to be balanced with the cost associated with the geometrically necessary regions with higher energy stacking sequences.
Reconstruction requires in-plane and possibly even out-of-plane atomic displacements, and thus introduces a local in-plane strain and an associated strain energy.
In other words the reconstruction is induced by interplanar interactions but opposed by intraplanar interactions.
The driving force for reconstruction can thus be expected to be larger in materials with large energy difference between different stacking sequences, and small in-plane stiffness (allowing for larger relaxations).

\begin{figure}
\centering
\includegraphics{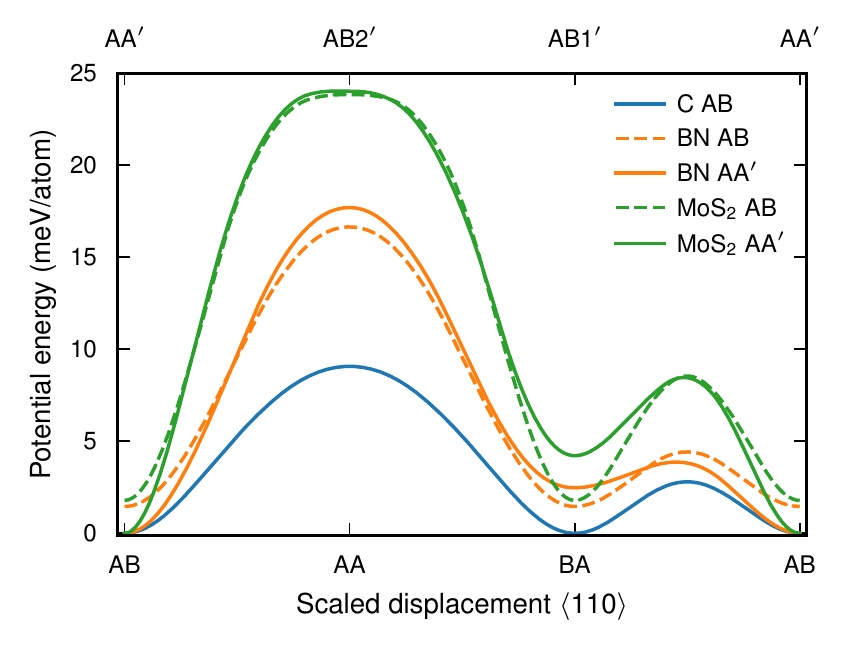}
\caption{
    Slip surfaces along the $\langle 110 \rangle$ direction.
    Solid lines indicate the energy surface associated with the respective ground state structure for each material (AB for C and AA$^\prime$ for \ce{BN} and \ce{MoS2}).
    Dashed lines indicate the slip surfaces based on the AB stacking for \ce{BN} and \ce{MoS2}.
    Energies between labeled structures are calculated for geometrically interpolated structures, corresponding to a translation of every other monolayer along the $\langle 110 \rangle$ direction.
    Here, BA refers to a symmetrically equivalent stacking of AB.
}
\label{fig:slip-surface}
\end{figure}

The slip surfaces (\autoref{fig:slip-surface}), which provide a picture of the energy landscape for in-plane displacements, show that \ce{MoS2} exhibits much larger energy differences between the different stackings than C and \ce{BN}.
Furthermore, the in-plane elastic constants (in-plane stiffness) are about 5 times larger in C and \ce{BN} compared to \ce{MoS2}.
Both of these effects contribute to a larger driving force for reconstruction in \ce{MoS2}, and in fact it is for \ce{MoS2} that one observes the most extended low energy stacking regions and the most narrow domain walls (\autoref{fig:moire-template-matching}).

The differences in  slip surface, reconstruction, order-parameter, and thermal conductivity of the moiré structures in the three materials considered here thus form a coherent picture.
Larger energy differences between the different stacking patterns along with smaller in-plane stiffness allow for more extensive reconstruction.
The latter leads to larger entropy in the stacking order parameter, which is in turn correlated with the \gls{ltc}.

\section{Conclusions}

The results and analysis provided in this study, building on earlier work for \ce{MoS2} \cite{Kim2021}, demonstrate that interlayer rotations in \gls{vdw} materials can be used for various different chemistries to control the through-plane \gls{ltc} while leaving the in-plane \gls{ltc} largely unchanged.
For all three materials considered here (C, \ce{BN}, and \ce{MoS2}) \emph{rotational disorder} in the form of stacks with random interlayer rotations leads to a very substantial reduction in the through-plane \gls{ltc} (\autoref{fig:kappa_stack}), resulting in a very large anisotropy between the through-plane and in-plane conductivities with ratios of about \num{1000} (C), \num{200} (\ce{BN}), and \num{300} (\ce{MoS2}) at \SI{300}{\kelvin}.
In all three cases, the through-plane \gls{ltc} is practically independent of temperature, indicating a glass-like conduction mechanism with minimal \gls{ltc} values of about \SI{1}{\watt\per\meter\per\kelvin} (C),  \SI{2}{\watt\per\meter\per\kelvin} (\ce{BN}), and  \SI{0.2}{\watt\per\meter\per\kelvin} (\ce{MoS2}).
The latter value can be compared with experimental data for \ce{MoS2} \cite{Kim2021}, which achieves an even lower level of \SI{57}{\milli\watt\per\meter\per\kelvin}.
The difference can likely be attributed to the presence of additional defects in the experimental samples and the strong sensitivity of the through-plane \gls{ltc} to the soft \gls{vdw}-mediated interlayer interactions.

The rotational disorder present in the stack systems causes the collapse of the transverse acoustic modes in the through-plane direction (\autoref{fig:phonon-dispersions}) and a reduction in the lifetimes of both the longitudinal and transverse acoustic modes (\autoref{sfig:LALO_lifetimes}).
This indicates that the mean free paths of the heat carrying modes become comparable to the interlayer spacing as expected in the glass limit.

Additional insight is provided by the dependence of the through-plane \gls{ltc} on the twist angle in \emph{rotationally ordered} moiré structures (\autoref{fig:ltc-vs-angle}).
For all three materials one observes a minimum in the through-plane \gls{ltc} between approximately 1 and \SI{3}{\degree}, which is most clearly pronounced in \ce{MoS2}.
In the latter case the minimal \gls{ltc} is moreover comparable to the value obtained in the stack system.
For C and \ce{BN}, on the other hand, there is still a notable gap between the minimal \gls{ltc} from moiré and stack structures, which we take as an indication that more than one layer must be rotated in order to reach the limiting value of the mean free path achieved in the stack structures.

We demonstrate that an entropy measure based on a simple order parameter for the stacking (dis)order, yields qualitative agreement with the angular dependence of the through-plane \gls{ltc} in the moiré structures.
This type of disorder is related to the moiré reconstruction which in turn can be related to the shape of the slip surface and the layer stiffness.
This strongly suggests that these quantities can be used as indicators for the efficacy of interlayer rotations as a means of reducing the through-plane \gls{ltc}.

Overall, the present results show that interlayer rotations can provide a chemistry agnostic approach to controlling the through-plane \gls{ltc} and the anisotropy ratio.

\section*{Contributions}
FE performed the \gls{ltc} calculations and construction of the disordered stacks.
EF performed the dispersion and entropy analysis and generated training structures.
CL helped in the analysis of the stacking sequences and slip surfaces.
ZF provided support and guidance in the use of \textsc{gpumd} and \gls{gk} methods and implemented the code for testing the interlayer interactions.
PE carried out the \gls{dft} calculations and trained and validated the \gls{nep} models.
FE, EF, and PE jointly wrote the paper.

\section*{Notes}

The authors declare no competing financial interest.
The \gls{nep} models, databases of the \gls{dft} calculations as well as associated scripts are available on Zenodo (https://doi.org/10.5281/zenodo.7811020).

\section*{Acknowledgments}

This work was funded by the Swedish Research Council (grant numbers 2018-06482, 2020-04935, 2021-05072), the Area of Advance Nano at Chalmers, and the Chalmers Initiative for Advancement of Neutron and Synchrotron Techniques.
The computations were enabled by resources provided by the National Academic Infrastructure for Supercomputing in Sweden (NAISS) and the Swedish National Infrastructure for Computing (SNIC) at C3SE, NSC, HPC2N, and PDC partially funded by the Swedish Research Council through grant agreements no. 2022-06725 and no. 2018-05973.
We thank Federico Grasselli for helpful discussions on the \gls{gk} formalism and Michele Simoncelli for insights into the Wigner formalism.

\appendix

\section*{Methods}
\label{sect:computational-details}

\subsection{Stacking sequences}
\label{sect:stackings}

Let us briefly recapitulate the different stacking sequences supported by the three materials of interest in this study.
In the case of \ce{BN} and \ce{MoS2} one can distinguish five different bulk stackings; AA, AB, AA$^\prime$, AB1$^\prime$ and AB2$^\prime$, as exemplified in \autoref{fig:moire-template-matching}a,f for \ce{BN}.\cite{Gilbert2019}
In the case of carbon, only the AA and AB stacking sequences are symmetrically unique, where the latter is also known as Bernal stacking.

In the AA and AA$^\prime$ stackings, all atoms have a neighbor directly above and below.
These stackings can also be classified as open (alternatively sparse or eclipsed) and the hexagonal structure is clearly apparent.
In the AA stacking B is on-top of B and N is positioned on-top of N.
In the AA$^\prime$ stacking on the other hand different atom types are stacked on-top of each other.

The AB, AB1$^\prime$, and AB2$^\prime$ stackings can be classified as closed (alternatively dense or staggered).
Due to the lack of an inversion center for \ce{BN} and \ce{MoS2} there are three variants.
The AB stacking can be thought of as B on-top of N with N and B placed in alternating hexagons.
The primed AB structures, AB1$^\prime$ and AB2$^\prime$, feature the same atom type on top of each other whereas the respective other type resides inside the hexagons.

The primed and unprimed structures cannot be related to each to other via a simple translation of one layer but are instead related by a 60$^{\circ}$ rotation of one layer in combination with a translation.
This results in two different types of slip surfaces (\autoref{sect:slip-surface}.
For carbon the AB stacking is energetically the most favorable.
For \ce{BN} and \ce{MoS2} the lowest energy structures AA$^\prime$ and AB are very close in energy \cite{Gilbert2019} where the energetic ordering is sensitive to the level of theory, including the choice of \gls{xc} functional.
Therefore, for \ce{BN} and \ce{MoS2} we consider both AA$^\prime$ and AB stackings throughout the paper.
While there is only one ground structure for each material the other stacking sequences appear in the reconstructed moiré structures due to geometric constraints.

\subsection{Construction of \texorpdfstring{\glspl{mlp}}{MLP}}

We employed the the second (NEP2) \cite{Fan22} and third (NEP3) generation \gls{nep} scheme \cite{FanWanYin22} to build \glspl{mlp} for C (NEP2), \ce{BN} (NEP3), and \ce{MoS2} (NEP3) using the \textsc{gpumd} package \cite{FanWeiVie2017, FanZenZha21, FanWanYin22}.
The \textsc{calorine} \cite{calorine} and \textsc{ase} \cite{Larsen2017} packages were used to construct the \glspl{nep}, handle atomic structures, and set up \gls{md} simulations.

The \gls{nep} model uses a multi-layer perceptron neural network architecture with a single hidden layer.
The radial part of the atomic environment descriptor is constructed from linear combinations of Chebyshev basis functions while the three-body angular part is similarly built from Legendre polynomials.
For the radial part cutoffs of \SI{8}{\angstrom}, \SI{8}{\angstrom} and \SI{7}{\angstrom} are used for \ce{C}, \ce{BN}, and \ce{MoS2}, respectively.
For the angular part cutoffs of \SI{3.5}{\angstrom}, \SI{4}{\angstrom}, and \SI{4}{\angstrom} are used for \ce{C}, \ce{BN} and \ce{MoS2}, respectively.
The hidden layer contains 50 neurons for all systems.

The \gls{nep} models were trained using a bootstrapping procedure in combination with active learning.
The initial training set included primitive structures of the different stackings both strained and unstrained as well as moiré structures up to moiré index 6 (corresponding to an angle of about \SI{5}{\degree}) both fully relaxed and with random displacements generated using the Monte Carlo rattling procedure implemented in the \textsc{hiphive} package \cite{EriFraErh19}.
Further structures were generated by \gls{md} simulations run at temperatures between \SI{100}{\kelvin} and \SI{900}{\kelvin} of both bulk and moiré structures, and added to the training set over a few iterations (\autoref{stab:models}).
For model validation see \autoref{sfig:parity-plot-virial-C_CX} to \autoref{sfig:parity-plot-virial-MoS2}.

Lastly, we also employed the \gls{bop} model for \ce{MoS2} \cite{liang_parametrization_2009, stewart_atomistic_2013} used in Ref.~\cite{Kim2021} for comparison with our \gls{nep} model.

\subsection{\texorpdfstring{\Gls{dft}}{DFT} calculations}
The energy, forces, and virials for the training structures were obtained via \gls{dft} calculations that were carried out using the projector augmented-wave method \cite{Blo94} as implemented in the Vienna ab-initio simulation package \cite{KreHaf93, KreFur96}.
The \gls{xc} contribution was represented using the \gls{cx} method \cite{DioRydSch04, BerHyl2014}.
For C we also carried out calculations using the PBE+D3(BJ) \cite{PerBurErn96, GriAntEhr10, GriEhrGoe11} and the \gls{scan} functionals \cite{SunRuzPer15}.
The Brillouin zone was sampled using $\Gamma$-centered grid with a linear $\vec{k}$-point spacing of about \SI{0.25}{\per\angstrom} and Gaussian smearing with a width of \SI{0.1}{\electronvolt}.
For the calculation of the forces a finer support grid was employed to improve their numerical accuracy.
All calculations were carried out using a plane-wave energy cutoff of \SI{520}{\electronvolt}.

\subsection{Thermal conductivity via \texorpdfstring{\gls{gk}}{GK}}
The \gls{gk} method  was used to calculate the \gls{ltc} as implemented in the \textsc{gpumd} package.
Specifically, the \gls{emd} method was employed and for each structure and temperature \num{100} independent production runs with a length of \SI{1}{\nano\second} were performed in the microcanonical ($NVE$) ensemble.
The simulations were equilibrated for \SI{100}{\pico\second} in the canonical ($NVT$) ensemble using the Langevin thermostat \cite{BusPar2007}.
The heat current was sampled every \SI{10}{\femto\second} and the running thermal conductivity was extracted at \SI{500}{\pico\second} using the Helfand-Einstein method \cite{Einstein1905, Helfand1960, Grasselli2021}.
The equilibrium lattice parameters for each temperature were found via isobaric-isothermal ($NPT$) \gls{md} simulations using stochastic velocity \cite{BusDonPar07} and cell rescaling \cite{BerBus20}.
A time step of \SI{1}{\femto\second} was used for all simulations.

The simulations of the stack systems were performed using a \numproduct{2x2x2} supercell for a total of \num{63984} atoms in the case of C and \ce{BN}, and a total of \num{95976} in the case of \ce{MoS2}.
For the moiré simulations the repetition was $N\times N \times 6$ (i.e., 12 monolayers) where $N$ varied depending on the index of the moiré cell so that the total number of atoms stayed above approximately \num{23000}.

\subsection{Mode projection}
\label{sect:mode-projection}

We employed phonon mode projection in order to analyze the phonons in the bulk and the stack systems from \gls{md} simulations \cite{SunSheAll2010, CarTogTan2017, RohLiLuoHen2022}.
The modes analyzed include the \gls{la}, \gls{lo}, \gls{ta}, and \gls{to} modes along $\Gamma \to$A.
For the stack system the bulk phonon modes were used, and although these are not exact harmonic eigenmodes of the system they are good approximations.
The autocorrelation function of the mode projected coordinate and velocity were fitted to damped harmonic oscillator functions in order to extract the frequencies and lifetimes of the \gls{la} and \gls{lo} modes \cite{FraSlaErhWah2021, FraRosEri22}.
\Gls{md} simulations were run in the microcanonical ensemble ($NVE$) for \SI{1}{ns} and results were averaged over about \num{50} independent runs.
These simulations were run using \num{40} monolayers for all systems.

\subsection{\texorpdfstring{\gls{ltc}}{LTC} from \texorpdfstring{\gls{pbte}}{PBTE}}

The phonon dispersions of the ideal bulk structures were also calculated using the \textsc{phonopy} package \cite{TogTan15}.
The \gls{ltc} for the ideal bulk structures was calculated using the direct solution of linearized phonon Boltzmann equation as implemented in the \textsc{phono3py} package \cite{phono3py}.
The force constants were obtained in a \numproduct{6x6x3} supercell and the \gls{ltc} was calculated  using a \numproduct{30x30x10} $\boldsymbol{q}$-point mesh.

\subsection{Stacking order parameter}
\label{sect:stacking-order-parameter}

To measure the stacking (or out-of-plane) disorder in moiré structures we introduce a simple atomic order parameter $\alpha_i$, which for atom $i$ is defined as
\begin{align}
    \alpha_i
    = \begin{cases}
    +1 - \sqrt{3} d_{ij} / a & \text{if $i$ and $j$ are the same species}\\
    -1 + \sqrt{3} d_{ij}/ a &  \text{else}
    \end{cases},
    \label{eq:order-parameter}
\end{align}
where
\begin{align*}
    d_{ij}^2 = \sqrt{\underset{j}{\min} \left[(x_i-x_j)^2 + (y_i-y_j)^2 \right]}
\end{align*}
is the shortest in-plane distance between atom $i$ and any atom $j$ in the neighboring layer.
For the bulk stacking sequences (\autoref{fig:moire-template-matching}a,f) one obtains $\alpha=-1, 0, 1$, whereas for the moiré structures $\alpha$ adopts continuous values between $-1$ and $+1$.
The stacking (or out-of-plane) disorder can the be estimated via the entropy as defined in Eq.~\eqref{eq:entropy}.

\subsection{Moiré structures}
The moiré structures were constructed according to the method described in Ref.~\citenum{lopesdossantos_graphene_2007}.
For all three materials moiré indices 1 through 11, 14, 22, 32, 45, 60, and 85 were included, corresponding to twist angles ranging from \SI{21.8}{\degree} to \SI{0.39}{\degree}.
For \ce{BN} and \ce{MoS2} two sets of moiré structures were constructed corresponding to the two distinct slip surfaces (\autoref{sect:slip-surface}).

\subsection{Stack structures}
The rotational disordered stacks were constructed by restricting the allowed in-plane strain of each layer to less than 1\%.
Each of the 10 layers contains approximately \num{400} primitive monolayer cells corresponding, e.g., to a \numproduct{20x20x5} AB stacked graphite supercell \cite{hermann_periodic_2012}.
The twist angles between the layers are 0, 1.44,  4.31,  7.15, 12.52, 17.48, 22.85, 25.05, 25.69, and \SI{28.56}{\degree}.
The stacks used here are slightly different from the ones used for simulations in Ref.~\citenum{Kim2021} but the results (\autoref{fig:kappa_stack}c) show this to have an indiscernible effect on the thermal conductivity.

\end{document}